\documentclass[prb]{revtex4}

\usepackage{graphicx}
\usepackage{amsmath}
\usepackage{amsfonts}

\begin{document}

\title{
Theory of Parity Violated Cooper Pairs
in Weakly Noncentrosymmetric Superconductors
}

\author{Satoshi Fujimoto} 
\affiliation{
Department of Physics,
Kyoto University, Kyoto 606-8502, Japan
}

\date{\today}

\begin{abstract}
We propose that in noncentrosymmetric superconductors 
with {\it weakly} asymmetric 
spin-orbit interaction the field-induced pair correlation between 
the spin-orbit split different bands
ignored in previous studies yields unique effects; i.e.  
the Pauli depairing effect is anisotropic in the momentum space, 
and as a result, magnetic fields
induce point-node-like anisotropic gap structure
of the quasiparticle energy
even for isotropic $s$-wave states, which seriously affects
thermodynamic quantities at low temperatures.  
Also, it is shown that when the magnitude of the spin-orbit interaction
is smaller than the superconducting gap,
the specific heat as a function of a magnetic field 
exhibit a two-gap-like behavior, even when there is only a single gap. 
These features characterize parity violated Cooper pairs
in {\it weakly} noncentrosymmetric systems.
We suggest the possible detection of these effects 
in the superconductor with weakly broken inversion symmetry
Y$_2$C$_3$. 
\end{abstract}


\maketitle
\section{Introduction}

One of the intriguing features of 
the recently discovered noncentrosymmetric superconductors 
\cite{bau,kim,onuki,uir,tog1}
is the realization of parity violated Cooper pairs, which leads 
various exotic phenomena.
\cite{ede1,ede2,gor,fri,sam1,ser2,yip,sam2,gor2,fei,kau,mine,fuji2,fuji3,fuji4,haya,haya2}
The investigations for unique effects characterizing parity violation 
have been mainly focused on the case with
strongly broken inversion symmetry (IS), where
the spin-orbit (SO) band splitting $E_{\rm SO}$ is enormous compared to
the superconducting (SC) gap $\Delta$,  
because it is commonly believed, and partly true, 
that effects due to broken IS are more prominent for stronger SO interaction.
However, in the present paper, we propose the possibility of novel phenomena
which characterize parity violated Cooper pairs 
in a unique way inherent in {\it weakly} noncentrosymmetric systems. 
These phenomena are essentially raised by the Zeeman effect on pairing states
in the SO split two bands.
In the weakly noncentrosymmetric case $E_{\rm SO}\sim \Delta $,
the Zeeman magnetic field induces substantial
pair correlation between the SO split different bands 
competing with the intra-band pairs, which is, in contrast, 
negligible in strongly noncentrosymmetric systems with $E_{\rm SO}\gg \Delta$. 
\cite{ede1,gor2,fei,kau,fuji2,fuji3,fuji4}
We demonstrate that the field-induced inter-band pair correlation
changes drastically low-energy properties of the SC state in the case 
with {\it weakly} broken IS, yielding the following unique effects; 
(i) the Pauli depairing effect is anisotropic in the momentum space, 
and as a result, magnetic fields induce the point-node-like structure of
the excitation gap even for isotropic s-wave states, which seriously affects
thermodynamic quantities, yielding distinct behaviors  
of them at low temperatures.
(ii) For $E_{\rm SO} < \Delta$, the specific heat as a function of
magnetic fields exhibits a two-gap like behavior even when
there is only a single SC gap.
These effects are associated with the momentum-dependent 
spin orientation of Cooper pairs which characterizes parity violation.
Thus, our results suggest a possible new direction of 
the experimental search for parity-violated Cooper pairs. 

As a matter of fact, 
our findings are relevant to the recent experimental studies on
the weakly noncentrosymmetric superconductor 
Y$_2$C$_3$.\cite{aki1,aki2,aku1,aku2}
This system has a cubic crystal structure with the space group symmetry
$I\bar{4}3d$ breaking inversion symmetry. 
Thus, the asymmetric SO interaction can be approximated 
by the Dresselhaus type interaction. 
The unique feature of this system is that 
the SO splitting is almost of the same order as 
the superconducting gap,\cite{ser,nis} 
and thus the situation considered in the present 
paper may be realized in this material.
In the last part of this paper, 
we shall compare our results with the experimental observations for 
this system, and discuss the possible
realization of the distinct phenomena 
characterizing parity violated
Cooper pairs in Y$_2$C$_3$.

Although the existence of the indispensable inter-band pairing correlation 
in addition to the intra-band pairs mentioned above plays an essentail role
in the above unique phenomena, it also brings about
some technical complexity of theoretical treatment, which hinders
the elucidation of properties of weakly noncentrosymmetric superconductors.
Moreover, the magnetic field induces the orbital depairing effect as well as
the Zeeman effect mentioned above.
To deal with these issues, we first analyze exactly 
effects of the Zeeman magnetic field
on the quasiparticle energy, neglecting the orbital depairing effect, 
and, afterward, to take into account the orbital effect, 
we develope the quasiclassical method which extends the classical works by Eilenberger\cite{eil} and Larkin-Ovchinnikov\cite{lo} to the case with 
both the intra-band and inter-band pairs.
A similar quasiclassical approach was considered before 
by Hayashi {\it et al.}\cite{haya}
Here, we obtain, for the first time, the explicit analytical solutions of the Eilenberger equations which encompass both the intra-band and inter-band pairs.
Using them, we discuss behaviors of thermodynamic quantities 
under applied magnetic field, 
in which the above-mentioned unique features 
characterizing parity violated Cooper pairs appear.

The organization of this paper is as follows.
In Sec.II, we investigate the Zeeman field effect 
in weakly noncentrosymmetric superconductor with 
the Dresselhaus type SO interaction, 
neglecting the orbital depairing effect, 
and demonstrate that in the case of $E_{\rm SO}\sim \Delta$ 
the Pauli depairing effect is anisotropic in the momentum space, leading
the point-node-like anisotropic structure of excitation energy gap.
This phenomenon should be 
important in the mixed state of type II superconductors.
Thus, in Sec.III, we develope the quasiclassical method, taking into account 
the orbital depairing effect in addition to
the above-mentioned anisotropic Pauli depairing effect.
Based upon this method, in Sec. IV,
we investigate the thermodynamic properties of the mixed state, and
elucidate how the above features characterizing the parity-violated Cooper
pairs appear in experimentally observable quantities.
It is demonstrated that for a sufficiently small SO interaction,
$E_{\rm SO} < \Delta$, 
the specific heat as a function of
magnetic fields exhibits a two-gap like behavior even when
there is only a single SC gap.
We shall also discuss the implication of our results for 
the recent experimental researches on 
the weakly noncentrosymmetric superconductor 
Y$_2$C$_3$.\cite{aki1,aki2,aku1,aku2}
Summary is given in Sec.V.

\section{Anisotropic Pauli depairing effect}

\begin{figure}[b]
\includegraphics*[width=10cm]{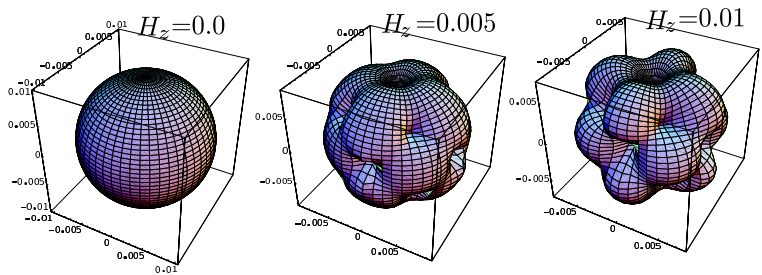}
\caption{\label{fig:fig1}  (Color online) The structure of 
the single particle excitation energy gap 
on the spherical Fermi surface of the band $\varepsilon_{k-}$
for $H_z=0.0$, $0.005$, $0.01$.}
\end{figure}

The absence of the inversion symmetry is characterized by the asymmetric
SO interaction,
\begin{eqnarray}
\mathcal{H}_{\rm SO}=\alpha\sum_{k,\sigma,\sigma'}\mbox{\boldmath 
$\mathcal{L}$}_0
(\mbox{\boldmath $k$})\cdot\mbox{\boldmath $\sigma$}_{\sigma\sigma'}
c^{\dagger}_{k\sigma}c_{k\sigma'},  
\label{soint}
\end{eqnarray}
where $c_{k\sigma}$ ($c^{\dagger}_{k\sigma}$) is the annihilation (creation)
operator of an electron with momentum $k$ and spin $\sigma$.
The components of $\mbox{\boldmath $\sigma$}=(\sigma^x,\sigma^y,\sigma^z)$ 
are the Pauli matrices.
Since we are concerned with the application to the cubic system Y$_2$C$_3$
with the space group symmetry $I\bar{4}3d$,
we assume the Dresselhaus interaction,
$\mbox{\boldmath $\mathcal{L}$}_0(\mbox{\boldmath $k$})=
(\mathcal{L}_{0x},\mathcal{L}_{0y},\mathcal{L}_{0z})=
(k_x(k_y^2-k_z^2),k_y(k_z^2-k_x^2),k_z(k_x^2-k_y^2))$.\cite{dres}

In the following, we consider the case of an $s$-wave state
with the isotropic SC gap $\Delta$, 
for which the features of weakly broken IS 
appear profoundly, as shown below.
We ignore the admixture with triplet pairs,
which is justified for $E_{\rm SO}/E_{\rm F}\ll 1$.
Our model Hamiltonian reads,
\begin{eqnarray}
\mathcal{H}=\mathcal{H}_{\rm BCS}+\mathcal{H}_{\rm SO},  \label{model}
\end{eqnarray}
\begin{eqnarray}
\mathcal{H}_{\rm BCS}=\sum_{k,\sigma}\varepsilon_kc^{\dagger}_{k\sigma}c_{k\sigma}
-\sum_k[\Delta c^{\dagger}_{k\uparrow}c^{\dagger}_{-k\downarrow}+
\Delta^{*} c_{-k\downarrow}c_{k\uparrow}].
\end{eqnarray}
In the following, the spherical Fermi surface,
$\varepsilon_k=k^2/(2m)-E_F$, is assumed for simplicity.
In this section, we concentrate on the effects of the Zeeman interaction, 
expressed by 
$-\sum_{k,\sigma\sigma'}\mu_{\rm B}\mbox{\boldmath $H$}
\cdot\mbox{\boldmath $\sigma$}_{\sigma\sigma'}c^{\dagger}_{k\sigma}c_{k\sigma'}$
with $\mbox{\boldmath $H$}=(0,0,H_z)$, 
leaving the analysis of the orbital depairing effect until the next sections.
When the Zeeman term is added to the Hamiltonian (\ref{model}),
the magnetic field induces the pairing between electrons
on the SO split different bands, which gives rise to
the Pauli depairing effect.
The important observation is that in the case of $E_{\rm SO}\sim \Delta$,
Cooper pairs with $\mbox{\boldmath $k$}$ for which the spin degeneracy
is not lifted by the SO interaction 
are more seriously
affected by the Pauli depairing effect than electron pairs
in the strongly SO split regions, leading to the anisotropic Pauli
depairing in the momentum space.
The anisotropic Zeeman effect in the strongly
noncentrosymmetric case has already been
discussed by several authors.\cite{sam1,mine}
The unique point in the weakly noncentrosymmetric case is that 
this effect 
drastically changes the structure of low-energy excitations
in the SC state, and yields the point-node-like structure of
the single-particle excitation gap even
in isotropic $s$-wave states.
To demonstrate this, we calculate the
single-particle excitation energy $E_{k\mu}$ ($\mu=1,2,3,4$) by diagonalizing 
the above mean field Hamiltonian 
(\ref{model}) with the Zeeman term expressed in the $4\times 4$ matrix form
in the space spanned by the basis 
$(c^{\dagger}_{k\uparrow},c_{-k\uparrow},c^{\dagger}_{k\downarrow},c_{-k\downarrow})$.
The explicit expressions for $E_{k\mu}$ are given in Appendix A.
In FIG.1, we depict the numerically calculated 
single-particle excitation energy gap $E_{\rm gap}$
for the band $\varepsilon_{k-}\equiv
\varepsilon_k-
|\alpha\mbox{\boldmath $\mathcal{L}$}_0(\mbox{\boldmath $k$}) 
-\mu_{\rm B}\mbox{\boldmath $H$}|$
in the case with $\Delta/E_{\rm F}=0.01$, and 
$\alpha/E_{\rm F}=0.03$.
Here $E_{\rm gap}$ is defined by
the magnitude of $E_{k3}$ for $k$ on the Fermi surface satisfying
$\varepsilon_{k-}=0$.
It is seen that
the point node-like structure develops,
as the magnetic field increases.
As a matter of fact, the excitation energy $E_{k\mu}$ is not truely
gapless, but the excitation energy gap $E_{\rm gap}$ is strongly anisotropic
with the structure similar to the point nodes, reflecting
the $k$-dependence of the SO term even when the superconducting gap
$\Delta$ is independent of $k$.
For instance, the excitation energy gaps $E_{\rm gap}$ for $\mbox{\boldmath $k$}_F \parallel (001)$, $\mbox{\boldmath $k$}_F \parallel (111)$
and $\mbox{\boldmath $k$}_F \parallel (100)$ (equivalent to
$\mbox{\boldmath $k$}_F \parallel (010)$)
are equal, and given by, 
\begin{eqnarray}
E_{\rm gap}^{(001)}=E_{\rm gap}^{(111)}=E_{\rm gap}^{(100)}
=\sqrt{(\mu_{\rm B}H_z)^2+\Delta^2}-\mu_{\rm B}H_z,
\label{gap1} 
\end{eqnarray}
while for $\mbox{\boldmath $k$}_F \parallel (110)$,
\begin{eqnarray}
E_{\rm gap}^{(110)}\approx \sqrt{2f(H_z)+\Delta^2-2
\sqrt{f(H_z)^2+\mu_{\rm B}^2H_z^2\Delta^2}}
\end{eqnarray}
with $f(H_z)=2\alpha^2k_F^6+\mu_{\rm B}^2H_z^2$.
It is easily checked that $E_{\rm gap}^{(110)}>E_{\rm gap}^{(001)}$.
The positions of the gap minima coincide approximately with the zero points of 
$\mbox{\boldmath $\mathcal{L}$}_0(\mbox{\boldmath $k$})$ at which
the SO splitting vanishes.
The quasi-particle excitations for $\mbox{\boldmath $k$}_F \parallel (001)$,
$(111)$, and $(100)$ behave like Dirac fermions with {\em mass gap} given by
eq.(\ref{gap1}).
It should be stressed that in the situation considered here 
the single-particle energy gap $E_{\rm gap}$ does not coincide
with the superconducting gap $\Delta$, which is $k$-independent.
A similar anisotropic structure of the excitation gap also appears
for the band $\varepsilon_{k+}\equiv
\varepsilon_k+
|\alpha\mbox{\boldmath $\mathcal{L}$}_0(\mbox{\boldmath $k$}) 
-\mu_{\rm B}\mbox{\boldmath $H$}|$.

Although we use the Dresselhaus interaction in the present calculation,
the point-node-like structure  
appears generally for any forms of 
$\mbox{\boldmath $\mathcal{L}$}_0(\mbox{\boldmath $k$})$
which possess zero points
on the Fermi surface.

\section{Quasiclassical approach for the mixed state in the case with
both inter-band and intra-band pairings}

The anisotroppic Pauli depairing effect discussed in the previous section
is particularly important in type II superconductors. 
Thus, in the following, 
we consider the orbital depairing effect as well as
the Pauli depairing effect 
on the basis of the quasiclassical analysis. 
For this purpose, we consider the Green functions
for quasiparticles defined on the SO split bands,
$G^{(\mu\nu)}(x,x')=-\langle T_{\tau}\psi_{\mu}(x)
\psi^{\dagger}_{\nu}(x')\rangle$, and
$F^{(\mu\nu)}(x,x')=-\langle T_{\tau}\psi^{\dagger}_{\mu}(x)
\psi^{\dagger}_{\nu}(x')\rangle$ where 
$\psi_{\mu}(x)$, $\psi^{\dagger}_{\mu}(x)$
are the field operators for quasiparticles in the $\mu$-band corresponding to
the energy in the normal state $\varepsilon_{k\mu}=
\varepsilon_k+\mu 
\alpha |\mbox{\boldmath $\mathcal{L}$}_0(\mbox{\boldmath $k$})|$ 
with  $\mu =\pm $.\cite{com}
As mentioned above,
the inter-band Green functions plays important roles
in addition to the intra-band Green functions.
Fourier transforming $G^{(\mu\nu)}$ and $F^{(\mu\nu)}$, 
we introduce the quasiclassical Green functions
defined by 
\begin{eqnarray}
\hat{\mathcal{G}}(\hat{k},\mbox{\boldmath $r$},\varepsilon_n)=
\left(
\begin{array}{cccc}
g^{(++)} & -f^{(++)} & g^{(+-)} & -f^{(+-)} \\
f^{(++)} & \bar{g}^{(++)} & f^{(-+)\dagger} & \bar{g}^{(-+)} \\
g^{(-+)} & -f^{(-+)} & g^{(--)} &  -f^{(--)} \\
f^{(+-)\dagger} & \bar{g}^{(+-)} & f^{(--)\dagger} & \bar{g}^{(--)}
\end{array}
\right),
\end{eqnarray}
with
$g^{(\mu\nu)}(\hat{k},\mbox{\boldmath $r$},\varepsilon_n)
=\int \frac{d\varepsilon_k}{\pi}G^{(\mu\nu)}
(\mbox{\boldmath $k$},\mbox{\boldmath $r$},\varepsilon_n)$, 
$f^{(\mu\nu)}(\hat{k},\mbox{\boldmath $r$},\varepsilon_n)
=\int \frac{d\varepsilon_k}{\pi}F^{(\mu\nu)}
(\mbox{\boldmath $k$},\mbox{\boldmath $r$},\varepsilon_n)$, and
$\bar{g}^{(\mu\nu)}=
g^{(\mu\nu)}(-k,\mbox{\boldmath $r$},-\varepsilon_n)$.
Here $\mbox{\boldmath $k$}$ 
is the momentum conjugate to the relative coordinate 
$\mbox{\boldmath $x$}-\mbox{\boldmath $x$}'$,
$\mbox{\boldmath $r$}$ is the center of mass coordinate, 
$\hat{k}$ is a unit vector 
parametrizing the direction of momentum $\mbox{\boldmath $k$}$, and
$\varepsilon_n$ is the fermionic Matsubara frequency.
Hereafter, matrices $\breve{A}$ represent $2\times 2$ matrices
defined in the space spanned by the basis 
$(\psi_{\mu},\psi^{\dagger}_{\nu})$ where 
$\mu=\nu=\pm$ or $\mu=-\nu=\pm$,
and matrices $\tilde{B}$ represent those
in the two-dimensional space spanned by the band indices $+,-$.
Using the standard method,\cite{eil,lo,vec} we find that $\hat{\mathcal{G}}$
satisfies the Eilenberger equation in the clean limit,
\begin{eqnarray}
i\mbox{\boldmath $v$}\cdot\frac{\partial}
{\partial \mbox{\boldmath $r$}}
\hat{\mathcal{G}}+[\omega\tau^z-\hat{M}+\hat{\Delta},\hat{\mathcal{G}}]=0,
\label{eieq}
\end{eqnarray}
where
$\omega=2i\varepsilon_n+
\mbox{\boldmath $v$}\cdot\frac{2e}{c}\mbox{\boldmath $A$}$
with  $\mbox{\boldmath $A$}$ 
a vector potential, and
$\tau^z=\tilde{1}\otimes\breve{\sigma}^z$. 
The $4\times4$ matrix $\hat{M}$ is defined by
$\hat{M}(\hat{k})=\tilde{\sigma}^z\otimes (\alpha\breve{L}(\hat{k}))$,
where $\breve{L}(\hat{k})=\breve{1}(|\mbox{\boldmath $\mathcal{L}$}(\hat{k})|+
|\mbox{\boldmath $\mathcal{L}$}(-\hat{k})|)/2
+\breve{\sigma}^z(|\mbox{\boldmath $\mathcal{L}$}(\hat{k})|-
|\mbox{\boldmath $\mathcal{L}$}(-\hat{k})|)/2$,
and
$\alpha|\mbox{\boldmath $\mathcal{L}$}(\hat{k})|=
|\alpha\mbox{\boldmath $\mathcal{L}$}_0(\hat{k})-\mu_{\rm B}
\mbox{\boldmath $H$}|$.
The matrix gap function $\hat{\Delta}(\hat{k},\mbox{\boldmath $r$})$
is 
\begin{eqnarray}
\hat{\Delta}(\hat{k},\mbox{\boldmath $r$})=\left(
\begin{array}{cc}
\breve{\Delta}_{+}(\hat{k},\mbox{\boldmath $r$}) 
& \breve{\Delta}_2(\hat{k},\mbox{\boldmath $r$}) \\
-\breve{\Delta}_{2}(\hat{k},\mbox{\boldmath $r$}) 
& \breve{\Delta}_{-}(\hat{k},\mbox{\boldmath $r$}) 
\end{array}
\right)
\end{eqnarray}
with $\breve{\Delta}_{\pm}(\hat{k},\mbox{\boldmath $r$})
=i\breve{\sigma}_y{\rm Re}\Delta_{\pm}(\hat{k},\mbox{\boldmath $r$})+
i\breve{\sigma}_x{\rm Im}\Delta_{\pm}(\hat{k},\mbox{\boldmath $r$})$,
$\breve{\Delta}_{2}(\hat{k},\mbox{\boldmath $r$})
=\breve{\sigma}_x{\rm Re}\Delta_{2}(\hat{k},\mbox{\boldmath $r$})-
\breve{\sigma}_y{\rm Im}\Delta_{2}(\hat{k},\mbox{\boldmath $r$})$,
$\Delta_{\pm}(\hat{k},\mbox{\boldmath $r$})
=\Delta(\mbox{\boldmath $r$})s_{\pm}(\hat{k})$,
$\Delta_{2}(\hat{k},\mbox{\boldmath $r$})
=\Delta(\mbox{\boldmath $r$})s_2(\hat{k})$,
and, 
\begin{eqnarray}
s_{\pm}(\hat{k})=-[\xi_{+}(\hat{k})\xi_{-}(-\hat{k})
+\xi_{-}(\hat{k})\xi_{+}(-\hat{k})]\eta_{\mp}(\hat{k}) 
\end{eqnarray}
\begin{eqnarray}
s_2(\hat{k})=\xi_{+}(\hat{k})\xi_{+}(-\hat{k})
-\xi_{-}(\hat{k})\xi_{-}(-\hat{k})
\end{eqnarray}
where 
$\xi_{\pm}(\hat{k})=\sqrt{
(1\pm\mathcal{L}_z(\hat{k})
|\mbox{\boldmath $\mathcal{L}$}(\hat{k})|)/2}$, and
$\eta_{\pm}(\hat{k})=-(\mathcal{L}_x\pm i\mathcal{L}_y)/
\sqrt{\mathcal{L}_x^2+\mathcal{L}_y^2}$.
Here, we omit the corrections to the Fermi velocity 
from the SO interaction, which are not important 
for $E_{\rm SO}/E_{\rm F}\ll 1$.
In the case with strong SO interaction $\alpha\gg \Delta$,
the contributions from the inter-band Green functions
to eq.(\ref{eieq}) are negligible, and the two bands are decoupled.
This simplified case was previously 
studied by several authors.\cite{haya,haya2}

\begin{figure}
\includegraphics*[width=7cm]{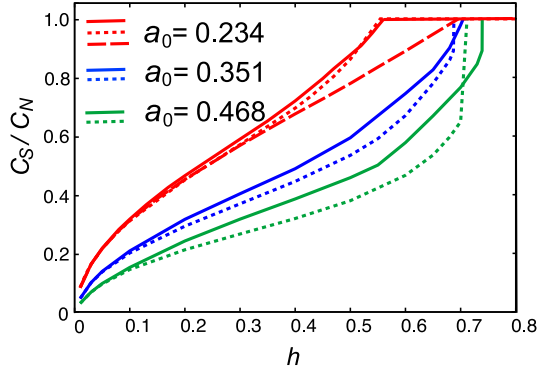}
\caption{\label{fig:fig2}  (Color online) Specific heat at $T=0.1T_c$ 
divided by the normal state value $C_N$ 
as a function of 
the normalized magnetic field $h=\mu_{\rm B}H_z/\Delta_0$ 
for $\alpha/\Delta_0=1.0$ (solid line), 
$\alpha/\Delta_0=0.0$ (dotted line),
and $\alpha/\Delta_0=10.0$ (dashed line). 
}
\end{figure}

In spite of the complications raised by the existence of both
the inter-band and intra-band Green functions, the analytical soutions
of (\ref{eieq}) based on the Pesch type approximation are possible.
In this approach, we assume the Abrikosov lattice solution
for $\Delta(\mbox{\boldmath $r$})$ and the uniform magnetic field
in the system, 
and replace the normal 
Green functions $g^{(\mu\nu)}$, $\bar{g}^{(\mu\nu)}$
with the spatial averages over a unit cell of the vortex lattice,
$\langle g^{(\mu\nu)}\rangle$, $\langle \bar{g}^{(\mu\nu)}\rangle$,
retaining only the spatial variation of $f^{(\mu\nu)}$ and 
$f^{(\mu\nu)\dagger}$.
Utilizing 
the normalization condition 
$\hat{\mathcal{G}}\cdot\hat{\mathcal{G}}=\mbox{\boldmath $1$}$
and the relations ${\rm tr}[\hat{\mathcal{G}}]=0$, $g^{(+-)*}=g^{(-+)}$,
which are derived from (\ref{eieq}),\cite{ref}
we find that the quasiclassical Green functions
are given by the solutions of the coupled algebraic equations,
\begin{eqnarray}
&&\sum_{\nu=\pm}|\langle g^{(\mu\nu)}\rangle |^2
-\sum_{\sigma,\nu=\pm}P(\tilde{\varepsilon}_n^{(\sigma\nu)})
|s_{\sigma\nu}\langle g^{(\sigma~\nu\sigma)}\rangle 
-s_{\sigma}\langle \bar{g}^{(\nu\sigma~\sigma)}\rangle \nonumber \\
&&-\nu\sigma s_2(\langle g^{(\sigma~-\nu\sigma)}\rangle
+\nu\langle \bar{g}^{(\nu\sigma~-\sigma)}\rangle)|^2 
=-1,  \quad (\mu=\pm),  \label{cae1}
\end{eqnarray}
\begin{eqnarray}
\langle g^{(+-)}\rangle=C
\sum_{\mu,\nu=\pm}
\nu [Y(\tilde{\varepsilon}_n^{(\mu\nu)})\langle g^{(\mu\mu)}\rangle 
-Y(\tilde{\varepsilon}_n^{(\nu\mu~\nu)})\langle\bar{g}^{(\mu\mu)}\rangle],
\label{cae2}
\end{eqnarray}
where 
$\tilde{\varepsilon}_n^{(\mu\nu)}=\varepsilon_n
+\frac{i\alpha\mu}{2}(|\mbox{\boldmath $\mathcal{L}$}|
-\nu |\mbox{\boldmath $\mathcal{L}'$}|)$, 
$\mbox{\boldmath $\mathcal{L}'$}=\mbox{\boldmath $\mathcal{L}$}(-\hat{k})$,
$P(\varepsilon)=\frac{1}{2}\partial Y(\varepsilon)/\partial \varepsilon$, 
$Y(\tilde{\varepsilon}_n)=
\sqrt{\pi}\Delta^2u_nW(2i u_n\tilde{\varepsilon}_n)$ with 
$\Delta^2=\langle \Delta^2(\mbox{\boldmath $r$})\rangle$, 
$W(z)=e^{-z^2}{\rm erfc}(-iz)$, and
$u_n=\Lambda ~{\rm sgn}\varepsilon_n/v\sin\theta$ 
with $\Lambda=\sqrt{\hbar c/2eH_z}$ and $\theta$ 
the polar angle of $\hat{k}$,
and $C=2is_2s_{+}\alpha |\mbox{\boldmath $\mathcal{L}'$}|/\tilde{D}$
with 
\begin{eqnarray}
\tilde{D}&=&4\alpha^2|\mbox{\boldmath $\mathcal{L}$}|
|\mbox{\boldmath $\mathcal{L}'$}|
+2i|s_{+}|^2\alpha(|\mbox{\boldmath $\mathcal{L}$}|+
|\mbox{\boldmath $\mathcal{L}'$}|)
\sum_{\mu=\pm}\mu Y(\tilde{\varepsilon}_n^{(\mu -)}) \nonumber \\
&&-2is_{2}^2\alpha(|\mbox{\boldmath $\mathcal{L}$}|
-|\mbox{\boldmath $\mathcal{L}'$}|)
\sum_{\mu=\pm}\mu Y(\tilde{\varepsilon}_n^{(\mu +)}).
\end{eqnarray}
Since the explicit expressions of the solutions are lengthy, we
will present them in Appendix B, and, instead, concentrate on
the discussion on the results obtained from them in the following section. 

\section{Specifc heat coefficient and density of states}

\begin{figure}
\includegraphics*[width=6.6cm]{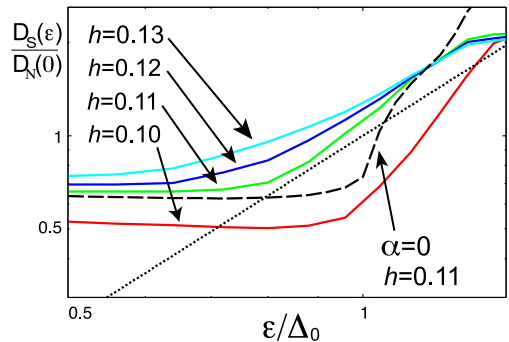}
\caption{\label{fig:fig3}  (Color online) 
Log-log plot of the density of states
versus energy for $\alpha/\Delta_0=1.0$.
The density of states for $\alpha=0.0$ and $h=0.11$
is also shown for comparison (broken).
The dotted line is $\varepsilon^2$.}
\end{figure}

Using the analytical solutions for $\hat{\mathcal{G}}$,
we calculate the specific heat and the density of states
of quasiparticles.
For this purpose, we determine the field-dependence and
the temperature-dependence of 
the spatially averaged gap function 
$\Delta(H_z,T)=\sqrt{\langle \Delta^2 (\mbox{\boldmath $r$})\rangle}$,
by solving the quasiclassical BCS gap equation,
\begin{eqnarray}
\langle \Delta^2(\mbox{\boldmath $r$})\rangle
=\lambda_0T\sum_n\sum_{\hat{k}}\langle\Delta(\mbox{\boldmath $r$})
f_{\uparrow\downarrow}(\hat{k},\mbox{\boldmath $r$},\varepsilon_n)\rangle,
\end{eqnarray}
\begin{eqnarray}
f_{\uparrow\downarrow}=\xi_{-}\bar{\xi}_{-}f^{(-+)}+\xi_{+}\bar{\xi}_{+}f^{(+-)}
-\xi_{+}\bar{\xi}_{-}f^{(++)}
-\xi_{-}\bar{\xi}_{+}f^{(--)} 
\end{eqnarray}
where $\bar{\xi}_{\alpha}\equiv\xi_{\alpha}(-\hat{k})$. 
$\lambda_0$ is a dimensionless coupling constant.
To compare our calculated results with the experimental observations
for Y$_2$C$_3$ later,
we tune the values of $\lambda_0$ and the cutoff for the frequency sum 
$\varepsilon_c$
so as to realize $T_c=18$ K. (e.g. $\lambda_0=0.2703$, $\varepsilon_c=700$ K.)
In this calculation, an important parameter is
$a_0\equiv\sqrt{\mu_{\rm B}\hbar c /(2 e \Delta_0)}/(\pi\xi_0)$,
which is an inverse of the coherence length $\xi_0$ normalized 
to be dimensionless. Here $\Delta_0$ is the gap function for $H_z=0$.
For Y$_2$C$_3$, $a_0\approx 0.234$.
The specific heat in the SC state is
\begin{eqnarray}
C_S=\int d\varepsilon
\frac{\varepsilon^2}{4T^2\cosh^2\frac{\varepsilon}{2T}}
D_S(\varepsilon),
\end{eqnarray}
where the density of states for quasiparticles $D_S(\varepsilon)$
is given by
$D_S(\varepsilon)=D_N(0)\sum_{\hat{k}}\sum_{\mu=\pm}{\rm Im}\langle 
g^{(\mu\mu)}(\hat{k},\mbox{\boldmath $r$},\varepsilon+i\delta)\rangle,$
with $D_N(0)$ the density of states in the normal state.
In FIG.\ref{fig:fig2}, we present the calculated results
of the specific heat as a function of
the normalized magnetic field $h=\mu_{\rm B}H_z/\Delta_0$
for the temperature $T/T_c=0.1$ and several sets of the parameters
$\alpha/\Delta_0$ and $a_0$.
It is found that in contrast to the case with strongly broken IS
where the SC state is quite robust against the Pauli depairing effect,
in weakly noncentrosymmetric systems
more low-energy excitations are induced by the magnetic field than
in the case with inversion symmetry $\alpha=0$.
We interpret the origin of this behavior
as the existence of the field-induced nodal excitations mentioned above.
The emergence of the point-node-like excitations
is more clearly observed in the energy-dependence of the
density of states for quasiparticles,
which is plotted in FIG.\ref{fig:fig3} for $\alpha/\Delta_0=1.0$.
As the magnetic field increases, the density of states exhibits
the power-law behavior $D_s(\varepsilon)\propto \varepsilon^2$.

\begin{figure}
\includegraphics*[width=7.2cm]{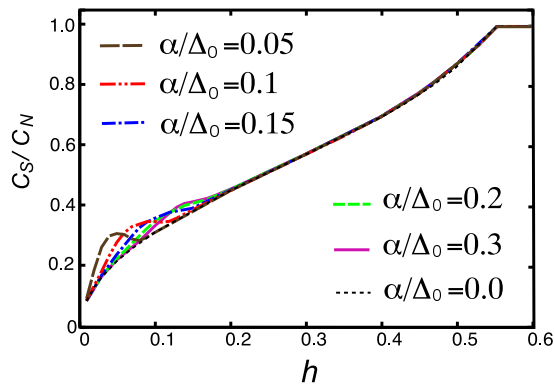}
\caption{\label{fig:fig4} (Color online) Specific heat as a function of 
the normalized magnetic field $h$ 
for small $\alpha/\Delta_0$. $a_0=0.234$. $T=0.1T_c$}
\end{figure}

When the SO coupling $\alpha$ is sufficiently smaller than $\Delta_0$,
another remarkable feature appears in the field-dependence of
the specific heat.
As shown in FIG.\ref{fig:fig4},
for the low magnetic fields $h\approx 0.1 \sim 0.2$,
a shoulder-like structure of $C_S(H_z)$ similar to a two-gap behavior appears,
though there is only a single SC gap $\Delta(\mbox{\boldmath $r$})$.
The origin of this behavior is understood as follows.
For $\alpha <\Delta_0$,
there are two different types of the Pauli depairing effect; 
i.e. (i) one due to 
the generation of the inter-band pairing correlation which, instead, suppresses
the intra-band pairing, and
(ii) the other caused by the asymmetric deformation of the Fermi surface.
The former is analogous to
the usual Pauli depairing effect in centrosymmetric superconductors.
The latter effect is inherent in noncentrosymmetric systems.
The crucial point is that although the former
exists for any finite magnetic fields, the latter is effective only 
for small fields $\mu_{\rm B}H_z<\alpha$, and is suppressed
for larger fields $\mu_{\rm B}H_z>\alpha$.
As a result, the character of the Pauli depairing effect
changes around $H_z\sim \alpha/\mu_{\rm B}$ in the case of
$\alpha <\Delta_0$, 
yielding the shoulder structure of the specific heat coefficient 
as demonstrated in FIG.\ref{fig:fig4}.
We would like to stress that this effect is caused by
the momentum dependent spin orientatin of Cooper pairs
characterizing parity violation.
Three remarks are in order. (1) For sufficiently small $\alpha$, e.g. 
$\alpha=0.05$, $C_S$ exhibits a hump rather than a shoulder, in marked 
contrast with conventional two-gap behaviors. 
(2) As $\alpha$ increases, the position of the shoulder shifts to
larger $h$ regions, though its structure becomes obscure since
the orbital depairing effect dominates 
for high magnetic fields.
(3) It should be cautioned that the Pesch approximation 
is not valid for  
magnetic fields much lower than $H_{c2}$.\cite{vec,pes}
However, the shoulder structure of $C_S$ at $h\sim 0.1$ 
stems from the Zeeman effect rather than orbital depairing effects.
Thus, the above results are applicable to
systems with a sufficiently large
value of the Ginzburg-Landau parameter $\kappa$, where
small magnetic fields can penetrate deeply into the SC regions.

We, now, discuss the implication of the above results 
for the experimental observations
of the weakly noncentrosymmetric superconductor Y$_2$C$_3$.  
This system is almost in the London limit with $\kappa > 10$.\cite{aki1,aku1} 
According to the recent LDA calculation, 
the averaged magnitude of the SO band splitting is 
roughly $\sim 0.01$ eV,\cite{nis} which is
of the same order as the SC gap $\sim 30$ K.\cite{aku1}
Thus, our analysis is applicable.
A remarkable experimental observation for Y$_2$C$_3$ is that 
the field-dependence of the specific heat 
exhibits a small shoulder structure for $H_z\sim 8\approx H_{c2}/3$ T 
at $T=2.6$ K.\cite{aku2}
Although this behavior was interpreted as the indication of
the existence of two SC gaps with different magnitudes,\cite{aki2,aku2}
it can be also, alternatively, explained by assuming the realization of
 the unique effect associated with parity violation
as demonstrated in FIG.~\ref{fig:fig4}.
A possible test for our scenario is to investigate
the field dependence of the nuclear spin relaxation rate,\cite{aki2}
which should exhibit a gap energy scale different from
that observed in the specific heat.

\section{Summary}

We have shown that in weakly noncentrosymmetric superconductors
the Pauli depairing effect is anisotropic in the momentum space,
inducing the point-node-like structure of 
the quasi-particle excitation energy gap 
even in isotropic $s$-wave states. 
This effect is caused by the competition between
the asymmetric SO interaction and the Zeeman magnetic field 
in the superconducting state, and yields unique low temperature behaviors
of thermodynamic quantities quite different from those of
conventional $s$-wave superconductors.
Also, by using the quasiclassical method, we have demonstrated that
the magnetic field dependence of the specific heat exhibits
a multi-gap-like structure for sufficiently small SO interaction
even when there is only a single gap.
These effects are associated with the momentum-dependent 
spin orientation of Cooper pairs which characterizes parity violation.
Thus, our results reveal
the unique aspects of parity violated Cooper pairs inherent
in weakly noncentrosymmetric systems.
We have also discussed that 
our findings may be relevant to the recent 
experimental observations for Y$_2$C$_3$.

\acknowledgments{}

The author would like to thank M. Sigrist, J. Akimitsu, S. Akutagawa,
Y. Nishikayama, H. Mukuda, A. Harada, and H. Ikeda for invaluable discussions.
The numerical calculations are performed on SX8 at YITP
in Kyoto University.
This work was partly supported by a Grant-in-Aid from the Ministry
of Education, Science, Sports and Culture, Japan.

\appendix
\section{Quasiparticle energy}
The single-electron energies of the Hamiltonian (\ref{model}) with the Zeeman term
$-\sum_{k,\sigma\sigma'}\mu_{\rm B}H_z
\sigma^z_{\sigma\sigma'}c^{\dagger}_{k\sigma}c_{k\sigma'}$
are given by the solutions of the following eigen value equation, 
\begin{eqnarray}
\left|
\begin{array}{cccc}
\varepsilon_k-\mu_{\rm B}H_z+\mathcal{L}_{0z}-x & 0 & 
\mathcal{L}_{0x}-i\mathcal{L}_{0y} & \Delta \\
0 & -\varepsilon_k+\mu_{\rm B}H_z+\mathcal{L}_{0z}-x & 
-\Delta^{*} & \mathcal{L}_{0x}+i\mathcal{L}_{0y}  \\
\mathcal{L}_{0x}+i\mathcal{L}_{0y}  & -\Delta &
\varepsilon_k+\mu_{\rm B}H_z-\mathcal{L}_{0z}-x & 0 \\
\Delta^{*} & \mathcal{L}_{0x}-i\mathcal{L}_{0y} & 0 &
-\varepsilon_k-\mu_{\rm B}H_z-\mathcal{L}_{0z}-x 
\end{array}
\right|=0.   \label{quartic}
\end{eqnarray}
Here the matrix is defined in the space spanned by
the basis 
$(c^{\dagger}_{k\uparrow},c_{-k\uparrow},c^{\dagger}_{k\downarrow},c_{-k\downarrow})$. 
The explicit solutions of (\ref{quartic}) are 
\begin{eqnarray}
E_{k1}=\frac{1}{2}\left(\sqrt{\tilde{\alpha}_{+}^{\frac{1}{3}}
+\tilde{\alpha}_{-}^{\frac{1}{3}}
-\frac{2}{3}X}+\sqrt{\frac{-2Y}
{\sqrt{\tilde{\alpha}_{+}^{\frac{1}{3}}+\tilde{\alpha}_{-}^{\frac{1}{3}}
-\frac{2}{3}X}}-\tilde{\alpha}_{+}^{\frac{1}{3}}-\tilde{\alpha}_{-}^{\frac{1}{3}}
-\frac{4}{3}X}\right), \label{dene1}
\end{eqnarray}
\begin{eqnarray}
E_{k2}=\frac{1}{2}\left(-\sqrt{\tilde{\alpha}_{+}^{\frac{1}{3}}
+\tilde{\alpha}_{-}^{\frac{1}{3}}
-\frac{2}{3}X}-\sqrt{\frac{2Y}
{\sqrt{\tilde{\alpha}_{+}^{\frac{1}{3}}+\tilde{\alpha}_{-}^{\frac{1}{3}}
-\frac{2}{3}X}}-\tilde{\alpha}_{+}^{\frac{1}{3}}-\tilde{\alpha}_{-}^{\frac{1}{3}}
-\frac{4}{3}X}\right), \label{dene2}
\end{eqnarray}
\begin{eqnarray}
E_{k3}=\frac{1}{2}\left(\sqrt{\tilde{\alpha}_{+}^{\frac{1}{3}}
+\tilde{\alpha}_{-}^{\frac{1}{3}}
-\frac{2}{3}X}-\sqrt{\frac{-2Y}
{\sqrt{\tilde{\alpha}_{+}^{\frac{1}{3}}+\tilde{\alpha}_{-}^{\frac{1}{3}}
-\frac{2}{3}X}}-\tilde{\alpha}_{+}^{\frac{1}{3}}-\tilde{\alpha}_{-}^{\frac{1}{3}}
-\frac{4}{3}X}\right), \label{dene3}
\end{eqnarray} 
\begin{eqnarray}
E_{k4}=\frac{1}{2}\left(-\sqrt{\tilde{\alpha}_{+}^{\frac{1}{3}}
+\tilde{\alpha}_{-}^{\frac{1}{3}}
-\frac{2}{3}X}+\sqrt{\frac{2Y}
{\sqrt{\tilde{\alpha}_{+}^{\frac{1}{3}}+\tilde{\alpha}_{-}^{\frac{1}{3}}
-\frac{2}{3}X}}-\tilde{\alpha}_{+}^{\frac{1}{3}}-\tilde{\alpha}_{-}^{\frac{1}{3}}
-\frac{4}{3}X}\right), \label{dene4}
\end{eqnarray} 
where 
$
\tilde{\alpha}_{\pm}^{\frac{1}{3}}=e^{\pm i\frac{4}{3}\pi}\alpha_{\pm}^{\frac{1}{3}},
$
and
\begin{eqnarray}
\alpha_{\pm}&=&\frac{1}{2}(\frac{1}{27}(2X^3+36XZ)-4XZ+Y^2 \nonumber \\
&\pm&
\sqrt{(\frac{1}{27}(2X^3+36XZ)-4XZ+Y^2)^2-\frac{4}{729}(12Z+X^2)^3})
\end{eqnarray}
with
\begin{eqnarray}
X=-2(\varepsilon_k^2+\alpha^2|\mbox{\boldmath $\mathcal{L}$}_0(k)|^2+
\mu_{\rm B}^2H_z^2+\Delta^2),
\end{eqnarray}
\begin{eqnarray}
Y=8\mu_{\rm B}H_z\alpha\mathcal{L}_{0z}\varepsilon_k, \label{Yc}
\end{eqnarray}
\begin{eqnarray}
Z&=&(\varepsilon_k^2-\alpha^2|\mbox{\boldmath $\mathcal{L}$}_0(k)|^2
-\mu_{\rm B}^2H_z^2)^2+\Delta^4+2\Delta^2\varepsilon_k^2
-4\mu_{\rm B}^2H_z^2\alpha^2\mathcal{L}_{0z}^2.
\end{eqnarray}
Since there is a term linear in $x$ in eq.(\ref{quartic}) 
with the coefficient given by (\ref{Yc})
for $H_z\neq 0$,
the particle-hole symmetry is broken by 
the Zeeman magnetic field.\cite{haya,ere}

\section{Solutions for quasiclassical Green functions}

In this appendix, we present the explicit solutions for the spatially averaged
quasiclassical 
Green functions satisfying eqs.(\ref{cae1}) and (\ref{cae2}),
which are derived from the Eilenberger equations (\ref{eieq})
combined with 
the normalization condition 
$\hat{\mathcal{G}}\cdot\hat{\mathcal{G}}=\mbox{\boldmath $1$}$.
Using the decoupling approximation for the spatial average,
$\langle g^{(\mu\nu)~2}\rangle\approx \langle g^{(\mu\nu)}\rangle^2$,
$\langle g^{(\mu\nu)}f^{(\kappa\lambda)}\Delta\rangle\approx
\langle g^{(\mu\nu)}\rangle \langle f^{(\kappa\lambda)}\Delta\rangle$, etc.,
we obtain the expressions for $\langle g^{(\mu\mu)}\rangle$,
\begin{eqnarray}
\langle g^{(++)}(p)\rangle&=&-i {\rm sgn}~\varepsilon_n[1-
P(\tilde{\varepsilon}_n^{(++)})|s_{+}(\hat{k})|^2\{1-r(p) \\ \nonumber
&-&2\alpha(|\mbox{\boldmath $\mathcal{L}'$}|
-|\mbox{\boldmath $\mathcal{L}$}|)s_2(\hat{k})(a(p)+b(p)r(p))\}^2 \\ \nonumber
&+&4\alpha^2|\mbox{\boldmath $\mathcal{L}'$}|^2|s_{+}(\hat{k})|^2
(a(p)+b(p)r(p))^2 \\ \nonumber 
&-&P(\tilde{\varepsilon}_n^{(--)})\{s_2(\hat{k})(1+r_2(-p))  \\ \nonumber 
&+&2\alpha(|\mbox{\boldmath $\mathcal{L}'$}|
+|\mbox{\boldmath $\mathcal{L}$}|)|s_{+}(\hat{k})|^2(a(p)+b(p)r(p))\}^2
]^{-\frac{1}{2}},
\end{eqnarray}
\begin{eqnarray}
\langle g^{(--)}(p)\rangle = -r(p)r_2(p)\langle g^{(++)}(p)\rangle,
\end{eqnarray}
where 
\begin{eqnarray}
a(p)=\frac{i s_2(\hat{k})}{\tilde{D}}[Y(1)-Y(3)+r_2(-p)(Y(2)-Y(3))],
\end{eqnarray}
\begin{eqnarray}
b(p)=-\frac{i s_2(\hat{k})}{\tilde{D}}[Y(1)-Y(4)+r_2(-p)(Y(2)-Y(4))],
\end{eqnarray}
\begin{eqnarray}
r_2(p)&=&\frac{1}{A_{+}}[A_{-}+8\alpha^3 (|\mbox{\boldmath $\mathcal{L}'$}|
-|\mbox{\boldmath $\mathcal{L}$}|)(|\mbox{\boldmath $\mathcal{L}$}|^2 \\ \nonumber
&-&|\mbox{\boldmath $\mathcal{L}'$}|^2)|s_{+}(\hat{k})|^2s_2(\hat{k})c_0^3Y(1)],
\end{eqnarray}
\begin{eqnarray}
A_{\pm}&=&Y(1)-2\alpha (|\mbox{\boldmath $\mathcal{L}'$}|
-|\mbox{\boldmath $\mathcal{L}$}|)s_2(\hat{k})c_0Y(1)  \\ \nonumber 
&-&4\alpha |\mbox{\boldmath $\mathcal{L}$}|s_2(\hat{k})c_0Y(3) \\ \nonumber 
&+&4\alpha^2 (|\mbox{\boldmath $\mathcal{L}'$}|  
+|\mbox{\boldmath $\mathcal{L}$}|)|s_{+}(\hat{k})|^2c_0^2
(Y(4)|\mbox{\boldmath $\mathcal{L}'$}|-Y(3)|\mbox{\boldmath $\mathcal{L}$}|)
\\ \nonumber 
&\pm&8\alpha^2(|\mbox{\boldmath $\mathcal{L}$}|^2
-|\mbox{\boldmath $\mathcal{L}'$}|^2)c_0|s_{+}(\hat{k})|^2 \\ \nonumber
&\times&
[c_{\pm}\{Y(1)-2\alpha|\mbox{\boldmath $\mathcal{L}$}|Y(3)c_0s_2(\hat{k}) 
\\ \nonumber
&-&\alpha (|\mbox{\boldmath $\mathcal{L}'$}|
-|\mbox{\boldmath $\mathcal{L}$}|)s_2(\hat{k})c_0Y(1) \} \\ \nonumber
&\pm& \alpha c_0^2 s_2(\hat{k})(Y(3)|\mbox{\boldmath $\mathcal{L}$}|
\mp Y(4)|\mbox{\boldmath $\mathcal{L}'$}|)]
\end{eqnarray}
\begin{eqnarray}
c_{+}=\frac{i s_2(\hat{k})}{\tilde{D}}[Y(2)-Y(1)+Y(4)-Y(3)]
\end{eqnarray}
\begin{eqnarray}
c_{-}=\frac{i s_2(\hat{k})}{\tilde{D}}[Y(4)+Y(3)-2Y(2)]
\end{eqnarray}
\begin{eqnarray}
c_{0}=\frac{i s_2(\hat{k})}{\tilde{D}}[Y(1)+Y(2)-2Y(3)]
\end{eqnarray}
\begin{eqnarray}
r(p)=\frac{1}{2R}[-S+{\rm sgn}~\varepsilon_n\sqrt{S^2-4RQ}],
\end{eqnarray}
\begin{eqnarray}
Q&=&Y(1)-2\alpha (|\mbox{\boldmath $\mathcal{L}'$}|
-|\mbox{\boldmath $\mathcal{L}$}|)s_2(\hat{k})Y(1)a(p) \\ \nonumber
&-&2\alpha |\mbox{\boldmath $\mathcal{L}'$}|Y(4)(1+r_2(p))s_2
(\hat{k})a(p)+a(p)^2h(p),
\end{eqnarray}
\begin{eqnarray}
R&=&-Y(1)-2\alpha (|\mbox{\boldmath $\mathcal{L}'$}|
-|\mbox{\boldmath $\mathcal{L}$}|)s_2(\hat{k})Y(1)b(p) \\ \nonumber
&+&2\alpha |\mbox{\boldmath $\mathcal{L}'$}|Y(4)(1+r_2(p))s_2
(\hat{k})b(p)+b(p)^2h(p),
\end{eqnarray}
\begin{eqnarray}
S&=&-2\alpha (|\mbox{\boldmath $\mathcal{L}'$}|
-|\mbox{\boldmath $\mathcal{L}$}|)s_2(\hat{k})Y(1)(a(p)+b(p))  \\ \nonumber 
&+&2\alpha |\mbox{\boldmath $\mathcal{L}'$}|Y(4)(1+r_2(p))s_2(\hat{k})a(p) 
\\ \nonumber 
&-&2\alpha |\mbox{\boldmath $\mathcal{L}$}|Y(3)(1+r_2(-p))s_2(\hat{k})b(p)
+2a(p)b(p)h(p),
\end{eqnarray}
with 
\begin{eqnarray}
h(p)&=&4\alpha^2(|\mbox{\boldmath $\mathcal{L}'$}|
+|\mbox{\boldmath $\mathcal{L}$}|)|s_{+}(\hat{k})|^2 \\ \nonumber 
&\times&(Y(4)|\mbox{\boldmath $\mathcal{L}'$}|
-Y(3)|\mbox{\boldmath $\mathcal{L}$}|).
\end{eqnarray}
Here $p=(\hat{k},\varepsilon_n)$, $Y(1)\equiv Y(\tilde{\varepsilon}_n^{(++)})$,
$Y(2)\equiv Y(\tilde{\varepsilon}_n^{(-+)})$,
$Y(3)\equiv Y(\tilde{\varepsilon}_n^{(--)})$, and
$Y(4)\equiv Y(\tilde{\varepsilon}_n^{(+-)})$.

Substituting the above expressions for $\langle g^{(\mu\mu)}\rangle$ into
eq. (\ref{cae2}), we obtain 
the off-diagonal components of the normal Green functions $\langle g^{(+-)}\rangle$,
and $\langle g^{(-+)}\rangle$.

Because of the existence of the Zeeman magnetic field, 
the particle-hole symmetry
does not hold, i.e. $\langle g^{(\mu\nu)}(-p)\rangle \neq -\langle
g^{(\mu\nu)}(p)\rangle$. Instead, we have,
\begin{eqnarray}
\langle g^{(++)}(-p)\rangle =r(p) \langle g^{(++)}(p)\rangle,
\end{eqnarray}
\begin{eqnarray}
\langle g^{(--)}(-p)\rangle =\frac{r_2(-p)}{r(p)r_2(p)} 
\langle g^{(--)}(p)\rangle,
\end{eqnarray}
\begin{eqnarray}
\langle g^{(+-)}(-p)\rangle =-\frac{|\mbox{\boldmath $\mathcal{L}$}(\hat{k})|}
{|\mbox{\boldmath $\mathcal{L}$}(-\hat{k})|}\langle g^{(+-)}(p)\rangle,
\end{eqnarray}
\begin{eqnarray}
\langle g^{(-+)}(-p)\rangle =-\frac{|\mbox{\boldmath $\mathcal{L}$}(\hat{k})|}
{|\mbox{\boldmath $\mathcal{L}$}(-\hat{k})|}\langle g^{(-+)}(p)\rangle.
\end{eqnarray}




\begin{thebibliography}{99}
\bibitem{com} In the derivation of the Gor'kov equations for 
$G^{(\mu\nu)}$ and $F^{(\mu\nu)}$,
we use the fact that the momentum which appears
in the unitary transformation from the basis of
$c_{k\sigma}$ ($c_{k\sigma}^{\dagger}$) to the basis of
$\psi_{\mu}(k)$ ($\psi_{\mu}^{\dagger}(k)$) commutes with
the center of mass coordinate $\mbox{\boldmath $r$}$
in the quasiclassical method.
\bibitem{ref} Because of the Zeeman effect, the particle-hole symmetry 
is broken, i.e. $g^{(\mu\nu)}+\bar{g}^{(\mu\nu)}\neq 0$.
This property of the Zeeman field is also noticed in
ref.\cite{haya} and \cite{ere}.
\end{thebibliography}
\end{document}